# A systematic study of four series of electron-doped rare earth manganates, $Ln_xCa_{1-x}MnO_3$ (Ln=La, Nd, Gd and Y) over the x=0.02-0.25 composition range

**L. Sudheendra, A.R. Raju and C.N.R. Rao***

*Chemistry and Physics of Materials Unit, Jawaharlal Nehru Centre for Advanced Scientific Research, Jakkur P.O., Bangalore, India-560064.*

*PACS: 75.30.Kz, 75.30.Vn, 75.50.Ee and 75.60.Ej*

Electrical and magnetic properties of four series of manganates $Ln_xCa_{1-x}MnO_3$ (Ln=La, Nd, Gd and Y) have been studied in the electron doped regime (x=0.02-0.25) in order to investigate the various inter-dependent phenomena such as ferromagnetism, phase separation and charge ordering. The general behavior of all the four series of manganates is similar, with some of the properties showing dependence on the average radius of the A-site cations, $\langle r_A \rangle$ and cation size disorder. Thus, all the compositions show increase in magnetization at 100-120 K ($T_M$) for $x<x_{max}$, the magnetization increasing with increasing x. The value of $x_{max}$ increases with decreasing $\langle r_A \rangle$, probably due to the increased phase separation induced by site disorder. This is also reflected in the larger width of the hysteresis loops at $T<T_M$ for small x or $\langle r_A \rangle$. In this regime, the electrical resistivity decreases with increasing x, but remains low and nearly constant $T>T_M$. The percolative nature of the conduction mechanism at $T<T_M$ is substantiated by the fit of the conductivity data to the scaling law, $\sigma \propto |x_c-x|^p$ where p is in the 2-4 range. When $x>x_{max}$, the materials become antiferromagnetic and charge-ordered at a temperature $T_{CA}$, accompanied by a marked increase in resistivity. The value of $T_{CA}$ increases with increase in $<r_A>$ and x (upto x=0.3). Thus, all the four series of manganates are characterized by a phase-separated regime between x=0.02 and 0.1-0.15 and an antiferromagnetic charge-ordered regime at x>0.1-0.15.

* For correspondence : e-mail : *cnrrao@jncasr.ac.in*

## 1. Introduction

The electron-doped regime of rare earth manganates of the general formula $Ln_xCa_{1-x}MnO_3$ (Ln=rare earth, x<0.5) have attracted considerable attention because of the unusual properties exhibited by the compositions in this regime [1,2]. The electron-doped manganates are different from the hole-doped manganates (x>0.5) in several ways, and electron-hole asymmetry in the manganate system has been a subject of discussion [2]. The electron-doped manganates are dominated by charge ordering and do not exhibit the ferromagnetic (FM) ground state at any composition. Even small doping in the A-site (x≤0.03) has marked effects on electronic properties of the manganates [3]. In the $Ln_xCa_{1-x}MnO_3$ system, compositions with x<0.03 exhibit G-type antiferromagnetism while those with x=0.1-0.15 shows C-type antiferromagnetism, often in admixture with G-type antiferromagnetism. Ferromagnetic clusters are present in the entire antiferromagnetic (AFM) regime [4,5]. Compositions in the x=0.03-0.1 range show a significant increase in magnetization around 100 K due to the presence of FM clusters in the AFM matrix. There is also a change in the resistivity behavior around 100 K and the x≤0.1 compositions exhibit low electrical resistivity above this temperature [6-8]. Magnetic susceptibility and resistivity data show some history dependence or irreversibility as well [5]. Charge ordering appears to manifest itself when x>0.1-0.15, but the manner of evolution of this phenomenon with change in temperature and composition has not been fully examined. It is generally believed that there is phase separation at low temperatures, with the size of the ferromagnetic domains increasing with the dopant concentration in the x=0.02-0.1 composition range [5,9]. Electrical conduction in this regime could be percolative.

In spite of several studies on the electron-doped regime of the $Ln_xCa_{1-x}MnO_3$ system reported in the literature, there are many aspects of these materials that we do not understand fully. Some important aspects of interest are the following: Is the ferromagnetic cluster or spin-glass regime (x<0.1-0.15) where the magnetization increases with x or electron concentration, sensitive to the average A-site cation radius, $\langle r_A \rangle$? Is the resistivity behavior metallic at temperatures beyond the spin-glass regime? In what way is the phase separation dependent on the composition and $\langle r_A \rangle$? Does size disorder play a role? Is the conduction percolative at low temperatures for x<0.1-0.15, and if so, is it affected by the A-site cation radius or site disorder? When does the charge-ordering manifest itself and what is the effect of the A-site cation radius on the charge-ordering transition? In order to contribute to the understanding of some of these issues, we have carried out systematic resistivity and magnetization measurements on $Ln_xCa_{1-x}MnO_3$ (Ln= La, Nd, Gd and Y) with x varying between 0.02 and 0.25, making sure that there are sufficient compositions in the x<0.1 and x>0.1 regimes. The latter regime (x=0.1-0.25) is relevant to examine the emergence of charge-ordering effects. Although measurements on some of these manganate compositions are reported in the literature, by Neumeier et al [3,4], Raveau and coworkers [5-9] and others, it was our considered view that careful measurements were necessary on a related series of manganates, covering a range of compositions from the pure G-AFM regime at very low x to the charge-ordered regime at x=0.1-0.25, and with different Ln substitutions, in order to fully understand the electron doped regime. The present study throws some light on the various phenomena such as phase separation, charge ordering and percolative conduction occurring in the

x=0.02-0.25 range of the electron-doped rare earth manganates and helps to adequately describe the characteristics of this fascinating regime.

## 2. Experimental

Polycrystalline powders of $Ln_xCa_{1-x}MnO_3$ (Ln = La, Nd,Gd and Y) compositions were prepared by the solid-state reaction of stoichiometric amounts of the rare earth acetate with $CaCO_3$ and $MnO_2$. The starting materials were ground and heated to 1000 $^0$C for 60 h with 3 intermediate grindings. Then the samples were reheated at 1200 $^0$C for 48 hrs with 2 intermediate grindings. The samples were then pelletized and heated to 1325 $^0$C for 36 h.  X-ray powder diffraction measurements were carried out using a Seifert 3000 diffractometer. All the manganate compositions had an orthorhombic structure, the lattice parameters varying with the Ln ion in the $Ln_xCa_{1-x}MnO_3$ (x=0.02-0.25) compositions. Thus, the lattice parameters decrease as we go from La to Y, but within a Ln series there is only a small variation with increase in x. The $\langle r_A \rangle$decreases with increase in x (except when Ln=La) in each series (Fig.1a). It is to be noted that the site disorder as measured by the variance $\sigma^2$ [10], also varies with x and the Ln ion. When Ln=La or Nd the $\sigma^2$ is relatively small, varying little with x, but when Ln=Gd or Y, $\sigma^2$ becomes appreciable increasing with x as shown in Fig.1b. The $\sigma^2$ is maximum and $\langle r_A \rangle$ smallest in the yttrium system Fig.1. The samples were analyzed for their elemental compositions by employing EDX analysis with quantitative ZAF correction software. Electrical resistivity measurements carried out by the standard four-probe method. Magnetization measurements were carried out at 4000 gauss using a LakeShore 7300 vibrating sample magnetometer.

## 3. Results and Discussion

In figures 2 and 3 we show the temperature variation of magnetization in several compositions of $Ln_xCa_{1-x}MnO_3$ with Ln= La, Nd, Gd and Y. The compositions with x≤ 0.1 in all the four series of manganates exhibit a marked increase in magnetization as in a ferromagnet around 100-120 K ($T_M$). These materials are, however, not real ferromagnets and show only small values of saturation magnetization even at 9000 gauss. At low x values (x≤0.1), the magnetization increases with x or the electron concentration and then decreases sharply. The maximum value of magnetization occurs at x=0.08, 0.08, 0.10 and 0.15 ($x_{max}$) for Ln = La, Nd, Gd and Y respectively. Accordingly, plots of $\mu_\beta$ versus x show maxima at increasing values of x as the average radius of the A-site cation, $\langle r_A \rangle$, decreases, the maximum value of $\mu_\beta$ being found when Ln=Gd (Fig.4). The increase in $x_{max}$ with decrease in $\langle r_A \rangle$ can arise from phase separation due to presence of significant FM fractions in the AFM matrix. The largest ferromagnetic fraction occurs around $x_{max}$ when the magnetization is maximum. In the composition range x>0.15 (x> $x_{max}$), the ferromagnetic fraction is small, with the concentration decreasing with increasing x. We would therefore expect phase separation to be prominent upto $x_{max}$. Considering that ferromagnetism itself would be favored by large $\langle r_A \rangle$, the observed trend in figure 4 can be taken to reflect an increase in the width of the phase separation regime. Interestingly, the $x_{max}$ is highest when Ln=Y. This is likely to be because of the large cation size disorder (Fig.1). This observation indirectly suggests that the phase separation in these manganate compositions is induced by size disorder, the separation regime increasing with $\sigma^2$.

After the magnetization attains a maximum value at $x_{max}$, we not only see a sudden drop in magnetization, but also evidence for a competition between ferromagnetism and antiferromagnetism. This competition gives rise to a peak in the magnetization-temperature curves (see figures 2 and 3). These peaks are reminiscent of the magnetization peaks found in charge-ordered systems such as $Nd_{0.5}Ca_{0.5}MnO_3$ and $Pr_{0.6}Ca_{0.4}MnO_3$ [1,2]. The temperature corresponding to the peak maximum ($T_{CA}$) increases with increasing x in the x=0.1-0.25 composition range in all the four series of manganates. The peak occurs at x~0.1 when Ln=La, at 0.15 for Nd and Gd and at 0.2 for Y. As shown in Fig. 5, $T_{CA}$ increases with $<r_A>$. It has been shown that the temperature corresponding to the magnetic susceptibility anomaly due to charge-ordering in electron-doped manganate compositions, increases upto x=0.3 and then decreases slightly in the x=0.3-0.5 composition range [2,11]. The effect of $<r_A>$ in negligible when $x \geq 0.3$.

The changes in the magnetization of $Ln_xCa_{1-x}MnO_3$ with composition and temperature are reflected in the electrical resistivities [6] as shown in figures 6 and 7. Thus, all the four series of manganates with $x<x_{max}$ show low resistivities from 300 K down to 100-120 K ($T_M$), independent of $<r_A>$. The activation energies for conduction are rather small (25-30meV) as reported by other workers [7,8]. In this low dopant concentration regime ($x<x_{max}$), the resistivity increases below $T_M$, the change possibly representing a semi-metal-insulator transition. The low-temperature resistivity (at $T<T_M$) increases with the decrease in electron concentration or x in this composition regime, paralleling the magnetization data in agreement with the earlier literature [6]. The resistivity also decreases with decrease in $<r_A>$ in the $Ln_xCa_{1-x}MnO_3$ series showing a minimum for Ln=Gd ($<r_A>$~1.179 Å) where the magnetization is maximum. Such a correspondence between the magnetization and the resistivity data is interesting and may have its origin in the phase separation resulting in percolative conduction. We shall examine this aspect later in the article.

The resistivity behavior drastically changes when $x>x_{max}$ in the four series of manganates. In this composition regime, the resistivity increases sharply as can be seen from figures 6 and 7. The increase in resistivity occurs around the same temperature as the peak in the magnetization-temperature curves ($T_{CA}$). The occurrence of a sharp change in resistivity at the same temperature as the magnetization peak suggests the occurrence of charge-ordering associated with antiferromagnetism. This transition temperature may, therefore, be considered to represent the onset of C-type antiferromagnetism.

The $x<x_{max}$ compositions in $Ln_xCa_{1-x}MnO_3$ exhibit magnetic hysteresis at $T<T_M$ (Fig.8). The hysteresis loops throw light on the nature of phase separation. In the multilayers of spin valve and permalloy materials [12-14] wherein the ferromagnetic layers are coupled to antiferromagnetic layers, the hysteresis loops reflect the extent of exchange coupling between the ferromagnetic and the antiferromagnetic layers. The coercivity of the ferromagnetic layer increases due the coupling with an antiferromagnetic layer. Since the phase-separated compositions of the manganates contain both the ferromagnetic and the antiferromagnetic regions, they would be expected to show similar behavior. We see from figure 8, that the hysteresis loops are broad at small x values ($x\leq0.04$) and the width decreases as x reaches $x_{max}$. When the system is subjected to magnetization reversal process, the interfacial spins between antiferromagnetic and ferromagnetic domains rotate with the ferromagnetic domain but experience an increased rotational drag due to the antiferromagnetic domains leading to broadening of the hysteresis loops. As the electron concentration or x increases, the ferromagnetic character (domain/cluster size) increases, thereby reducing the drag substantially and causing a decrease in the width as expected. It must be recalled that the ferromagnetic fraction reaches a maximum at $x_{max}$. The width of the hysteresis loop for a given x value increases with the decrease in the radius of the Ln (or $\langle r_A \rangle$), reflecting the effect of phase separation.

The ferromagnetic hysteresis loops are symmetric, indicating two equivalent directions of magnetization. On the other hand, when a ferromagnetic/antiferromagnetic material is cooled in an external magnetic field (as in this case for obtaining M Vs T plots) below the Neel temperature (< 100 K), the loop is shifted from zero due to the exchange bias effect. The small differences in the coercive fields (Table.1) seen for the phase-separated samples may be due to the pinning of a small fraction of ferromagnetic interfacial spins to the antiferromagnetic domains. The pinned spins do not rotate in an external field as they are coupled leading to exchange biasing.

Magnetization reversal in a ferromagnetic system occurs through rotation of spins, and via domain nucleation and growth. An examination of the hysteresis loops shows that magnetization drops from a maximum value to zero sharply compared to the increase from zero to the maximum value. This suggests that for $0.02<x\leq0.06$, one mechanism dominates over the other in a particular region of magnetization reversal process, bringing about an asymmetry in the hysteresis loops. Thus, it is much harder to magnetize the $0.02<x\leq0.06$ compositions wherein the ferromagnetic clusters are embedded in a antiferromagnetic matrix.

It was mentioned earlier that the conducting mechanism in the phase-separated regime maybe percolative. We have employed the scaling law, $\sigma \propto | x_c\text{-}x |^p$, to treat the resistivity data in the x=0.02-0.1 composition regime. We show $\ln\sigma$ -lnx plots at 50 K and 20 K from the series of manganates studied by us in Fig.9. The value of the exponent p is between 2.1 and 2.8 at 50 K and between 2.8 and 4.3 at 20 K. Percolative conduction becomes less dominant as the temperature approaches $T_M$ or x reaches $x_{max}$ (0.1-0.15). This is understandable since antiferromagnetism is the main interaction when $x>x_{max}$ and $T>T_M$, leading to lesser phase separation and a more homogeneous antiferromagnetic phase.

## 4. Conclusions

The present study of the four series of electron-doped manganates, $Ln_xCa_{1-x}MnO_3$ (Ln=La, Nd, Gd and Y), over a wide range of compositions (x=0.02-0.25) has been useful in understanding the evolution of various phenomena. Thus, these materials which show ferromagnetic-like behavior at $T<T_M$ upto a value of x ($x_{max}$), become antiferromagnetic with charge ordering at $T_{CA}$ for $x>x_{max}$. The values of $x_{max}$ and $T_{CA}$ depend on the average radius of A-site cation ($\langle r_A \rangle$), the former being related to the phase separation and site disorder ($\sigma^2$). Phase separation is favored at low temperatures ($T<T_M$) by small x ($<x_{max}$) and small $\langle r_A \rangle$ (or large $\sigma^2$). The AFM CO transition temperature, $T_{CA}$, on the other hand, increases with increasing x and $\langle r_A \rangle$ and does not vary significantly with $\langle r_A \rangle$ when $x\geq 0.3$. All these materials show low resistivity at $T>T_M$ when $x\leq x_{max}$, but show a sharp increase in resistivity at $T_{CA}$ when $x>x_{max}$, due to charge-ordering. When $x<x_{max}$ and $T<T_M$, conduction appears to be percolative. Since the above features are found in all the four series of manganates covering a wide range of $<r_A>$, they can be taken to be intrinsic to the electron doped compositions of the rare earth manganates. Thus, the present study clearly identifies the phase-separated regime with percolative conduction ($x=0.02-x_{max}$) and the charge-ordered antiferromagnetic regime with $x_{max}$ in the ~0.1-0.15 range. The percolative regime increases with size disorder or decrease in $<r_A>$.

**Table. 1** Coercivity ($H_c$ and $-H_c$ in gauss) of $Ln_xCa_{1-x}MnO_3$ obtained from hysteresis measurement at 50 K.

| **x** | **La** | | **Nd** | | **Gd** | | **Y** | |
|---|---|---|---|---|---|---|---|---|
| | $H_c$ | $-H_c$ | $H_c$ | $-H_c$ | $H_c$ | $-H_c$ | $H_c$ | $-H_c$ |
| 0.02 | 966 | 849 | 454 | 340 | 508 | 395 | 429 | 338 |
| 0.04 | 706 | 604 | 454 | 449 | 580 | 572 | 865 | 589 |
| 0.06 | 300 | 203 | 215 | 124 | 373 | 279 | 579 | 473 |
| 0.08 | 94 | 100 | 114 | 32 | 154 | 62 | 344 | 247 |
| 0.10 | - | - | 870 | 776 | 154 | 62 | 322 | 220 |
| 0.15 | - | - | - | - | - | - | 310 | 208 |

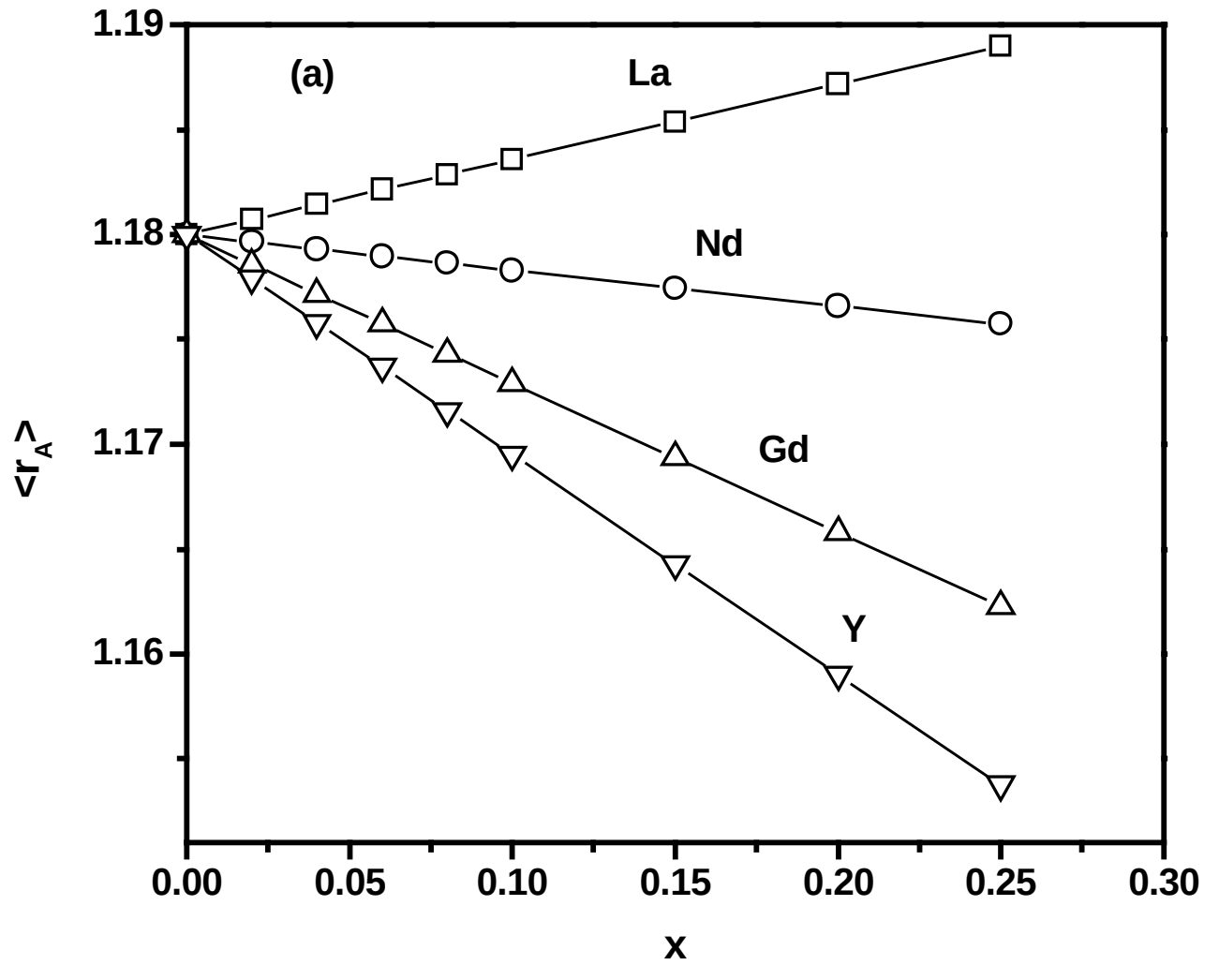

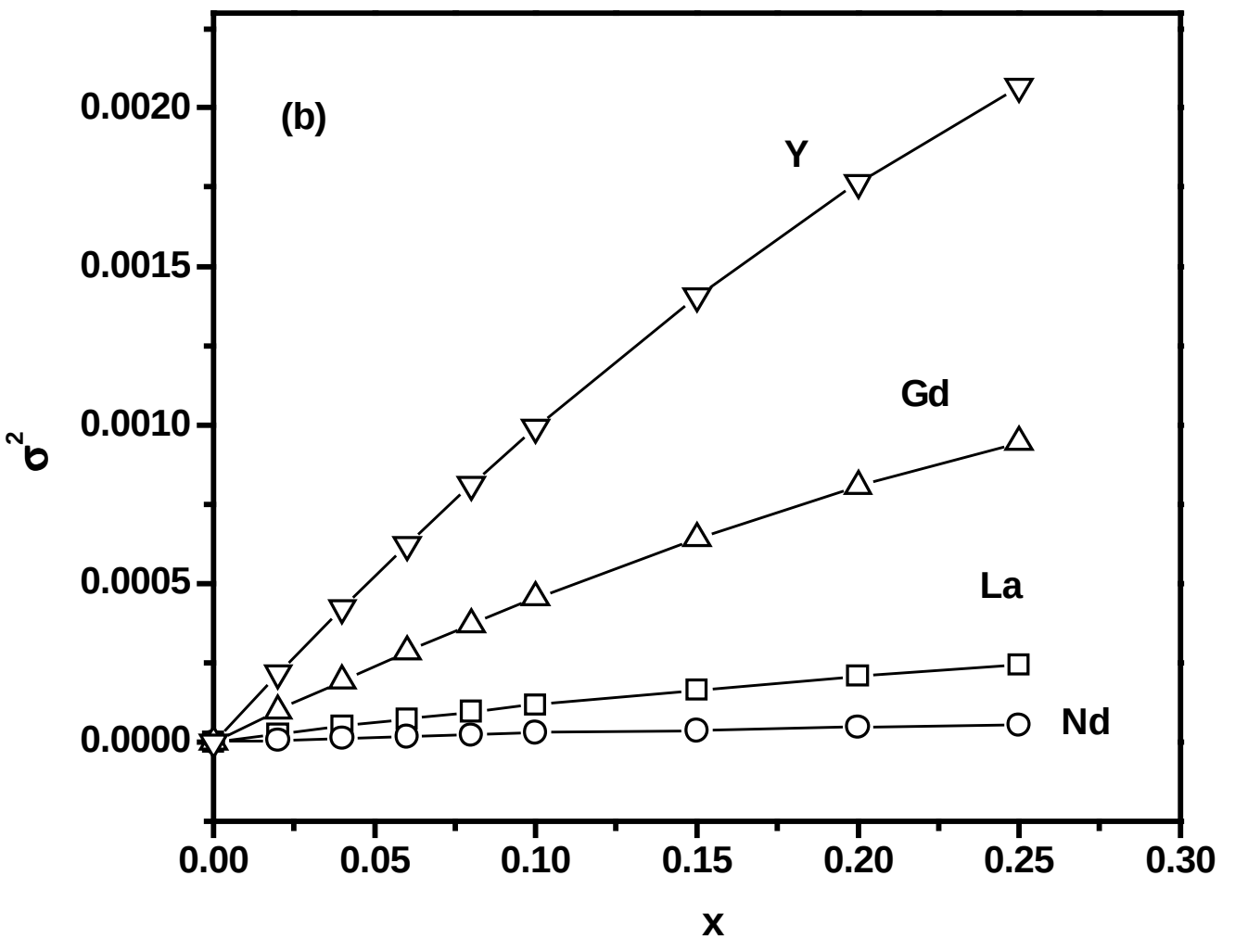

Fig.1 Variation of (a) $\langle r_A \rangle$ and (b) $\sigma^2$ with x in $Ln_xCa_{1-x}MnO_3$

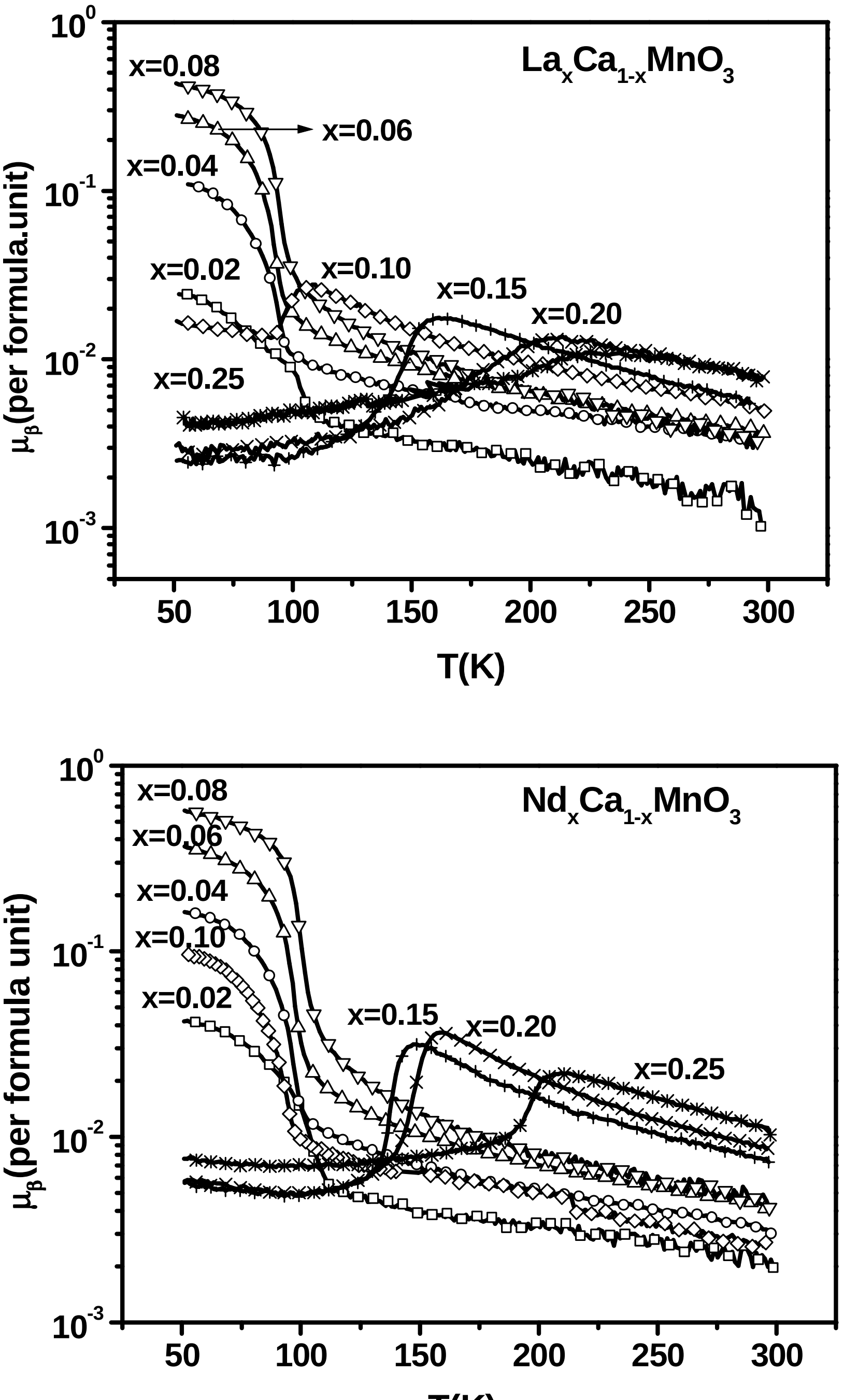

Fig.2 Temperature variation of magnetization of $Ln_xCa_{1-x}MnO_3$ with x=0.02-0.25: (a) Ln=La, (b) Ln=Nd.

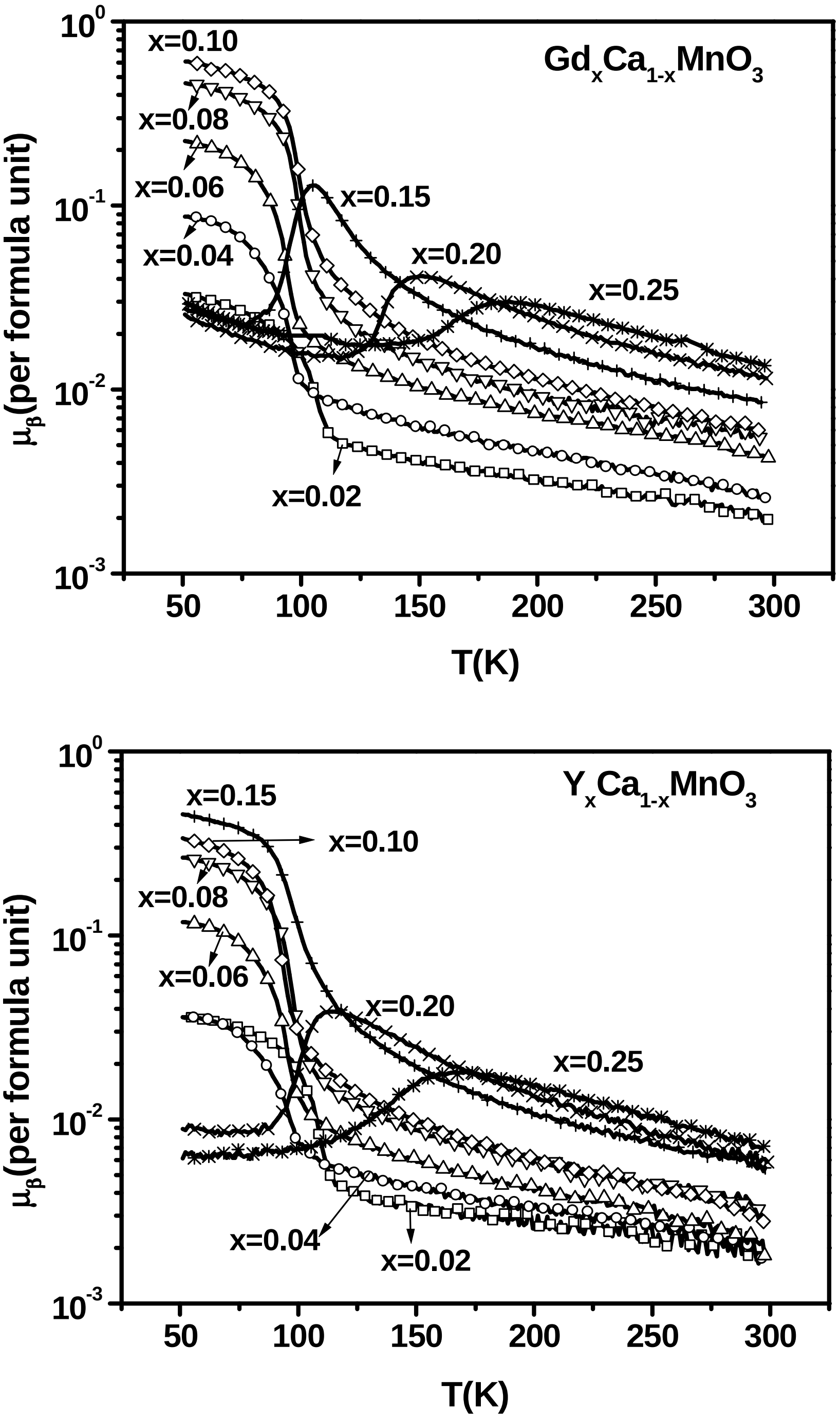

Fig.3 Temperature variation of magnetization of $Ln_xCa_{1-x}MnO_3$ with x=0.02-0.25: (a) Ln=Gd, (b) Ln=Y.

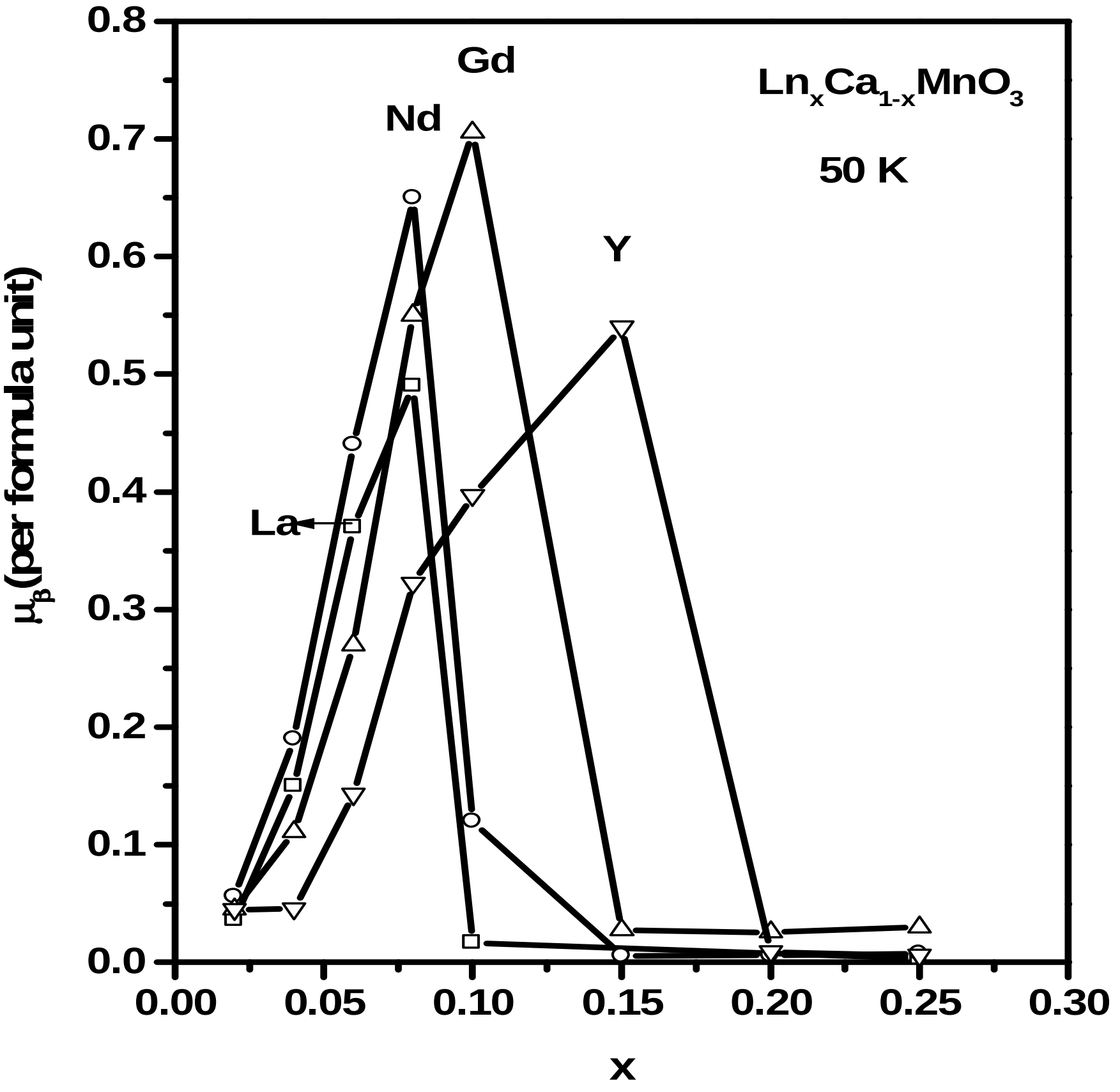

Fig.4 Variation of $\mu_\beta$ with composition at 50 K for $Ln_xCa_{1-x}MnO_3$.

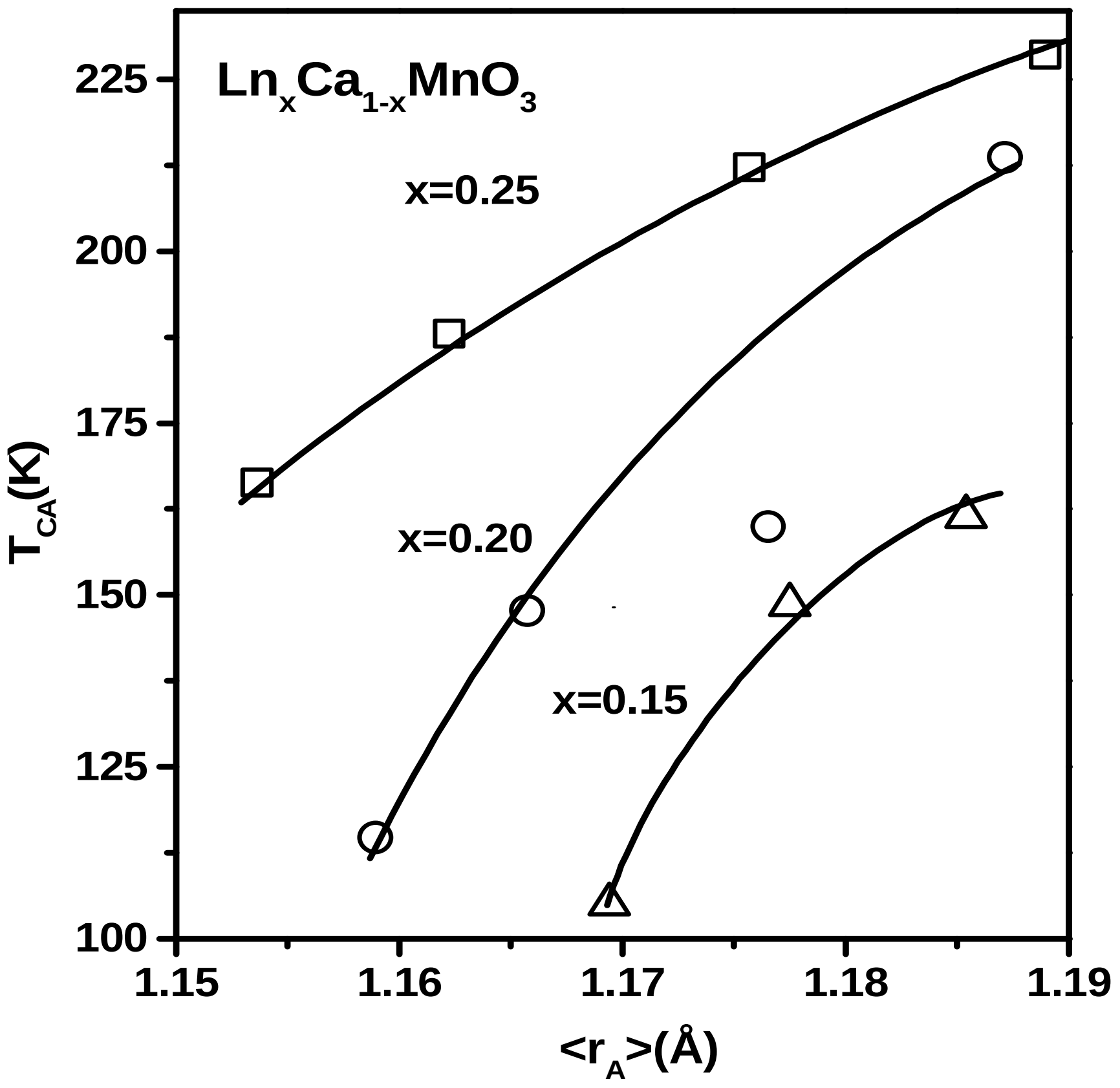

Fig.5 Variation of the antiferromagnetic charge ordering transition temperature, $T_{CA}$ with the average A-site cation radius ($<r_A>$) in $Ln_xCa_{1-x}MnO_3$ for different values of x ($x>x_{max}$).

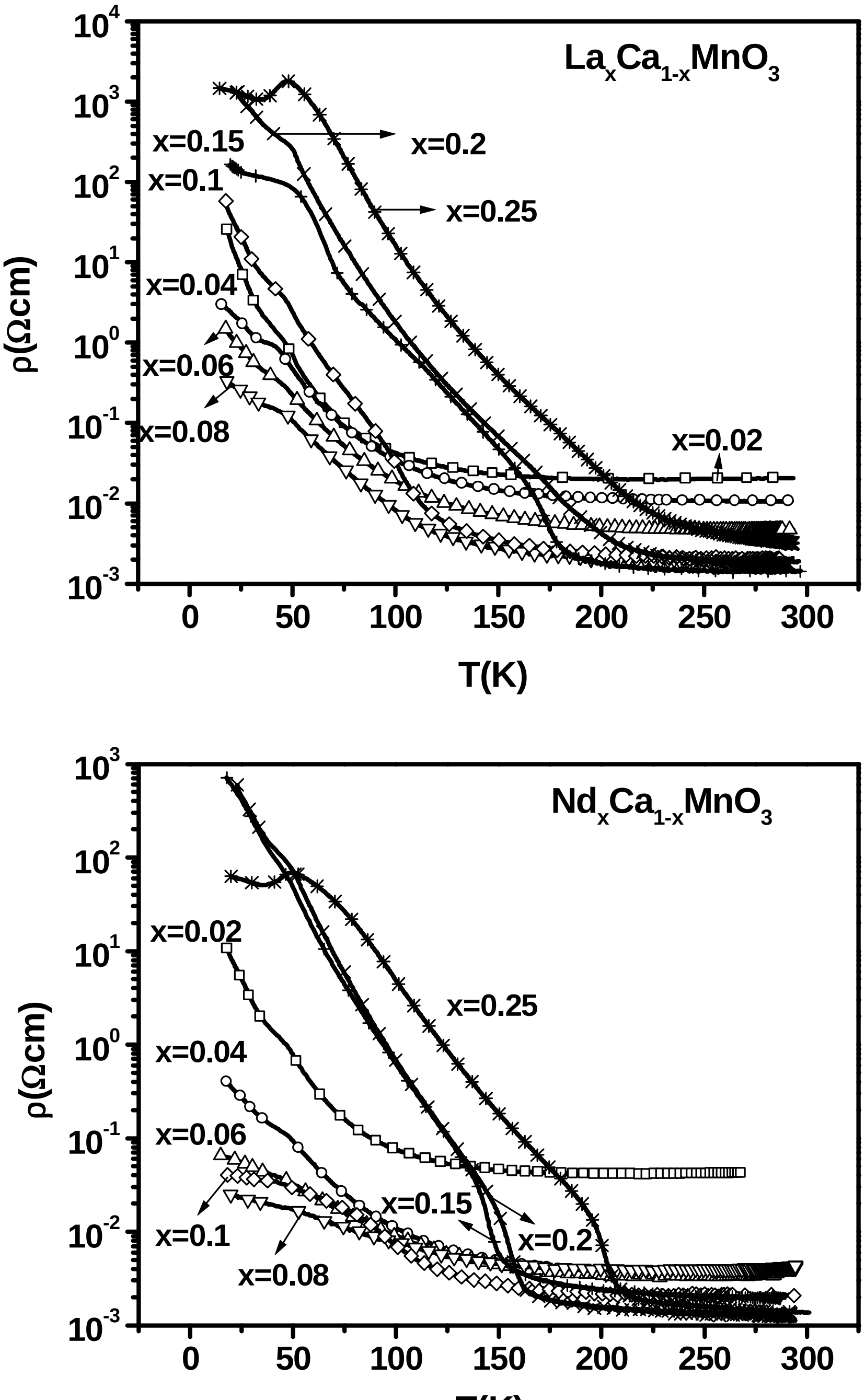

Fig.6 Temperature variation of the electrical resistivity of $Ln_xCa_{1-x}MnO_3$ with x=0.02-0.25: (a) Ln=La, (b) Ln=Nd.

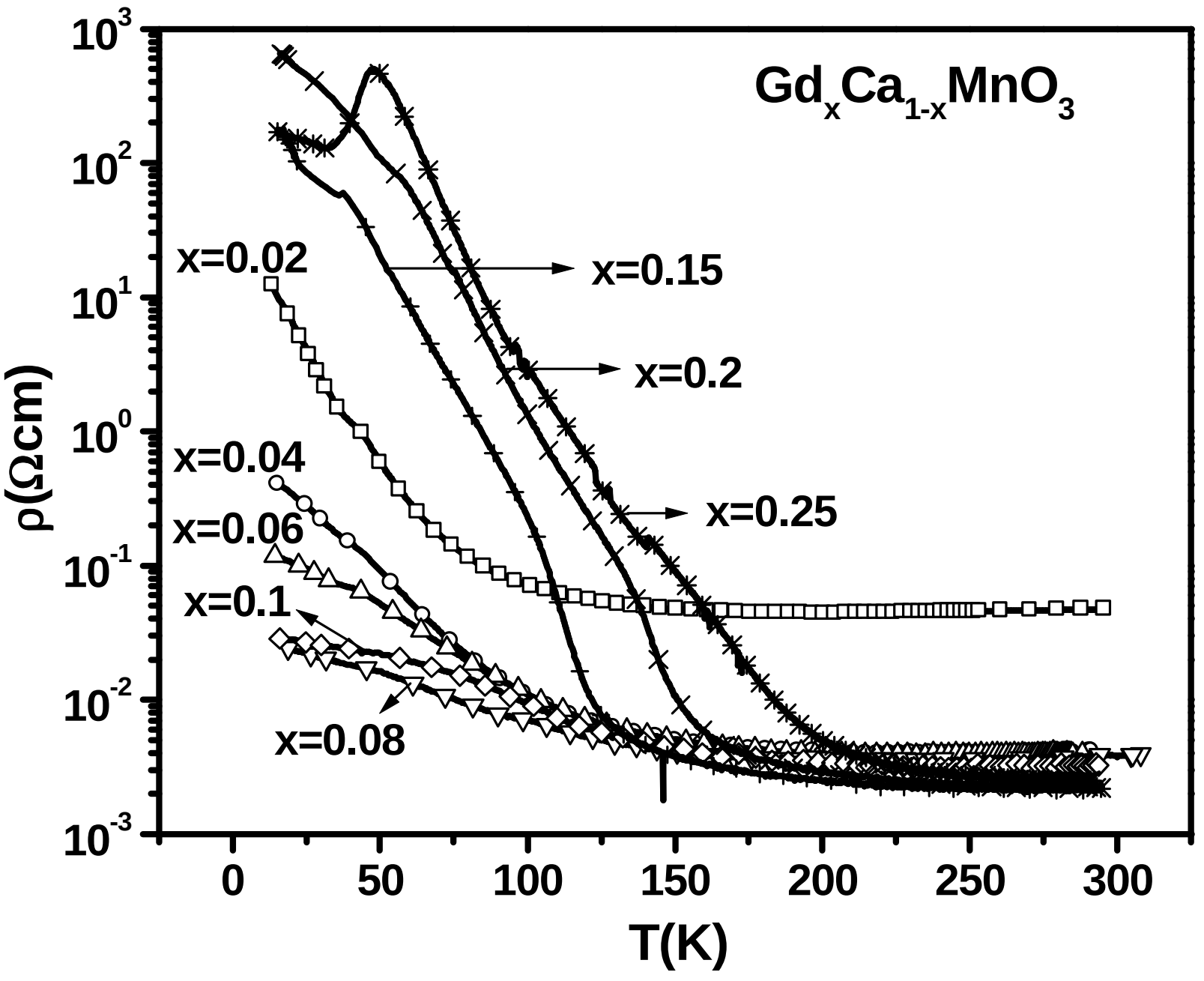

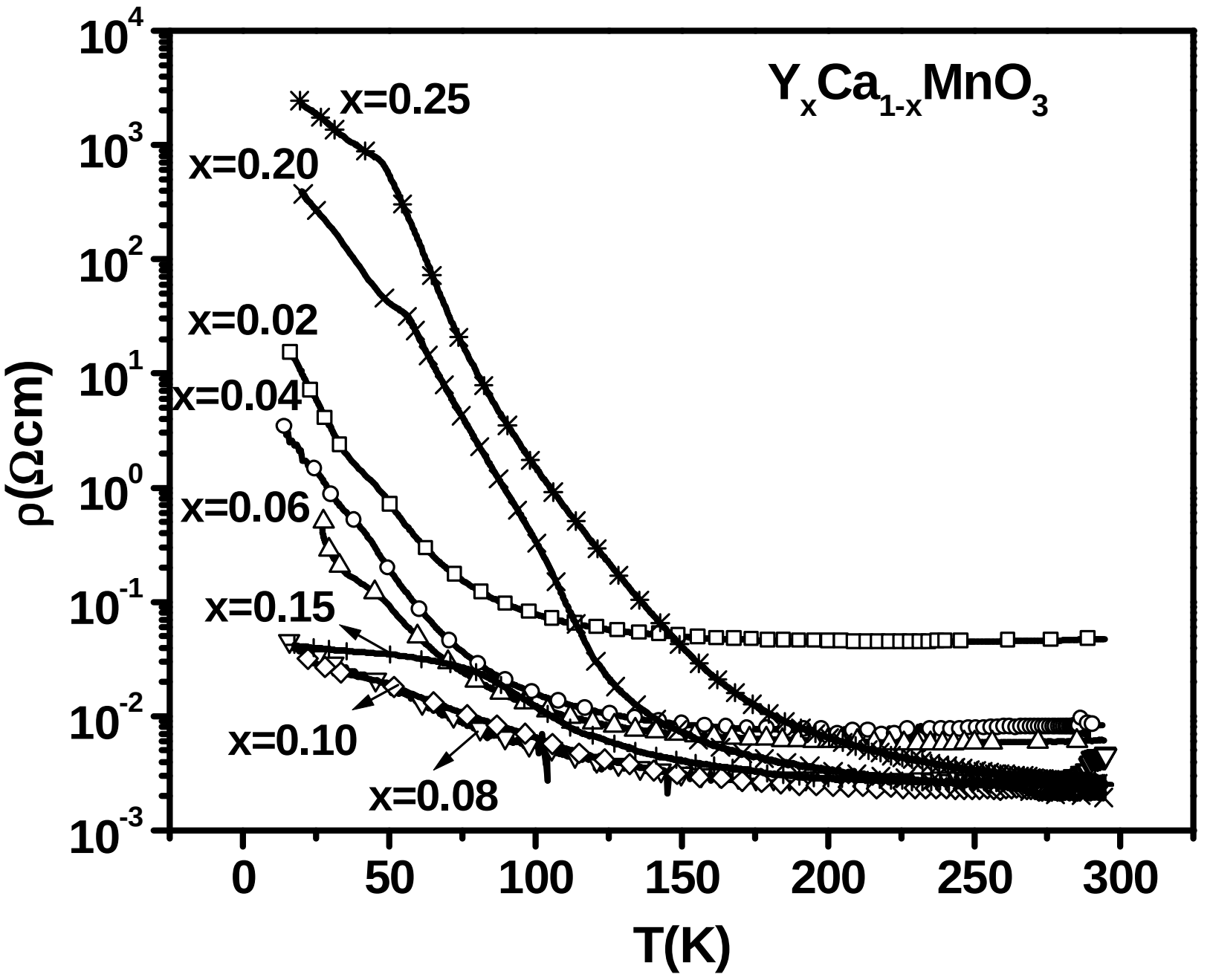

Fig.7 Temperature variation of the electrical resistivity of $Ln_xCa_{1-x}MnO_3$ with x=0.02-0.25: (a) Ln=Gd, (b) Ln=Y.

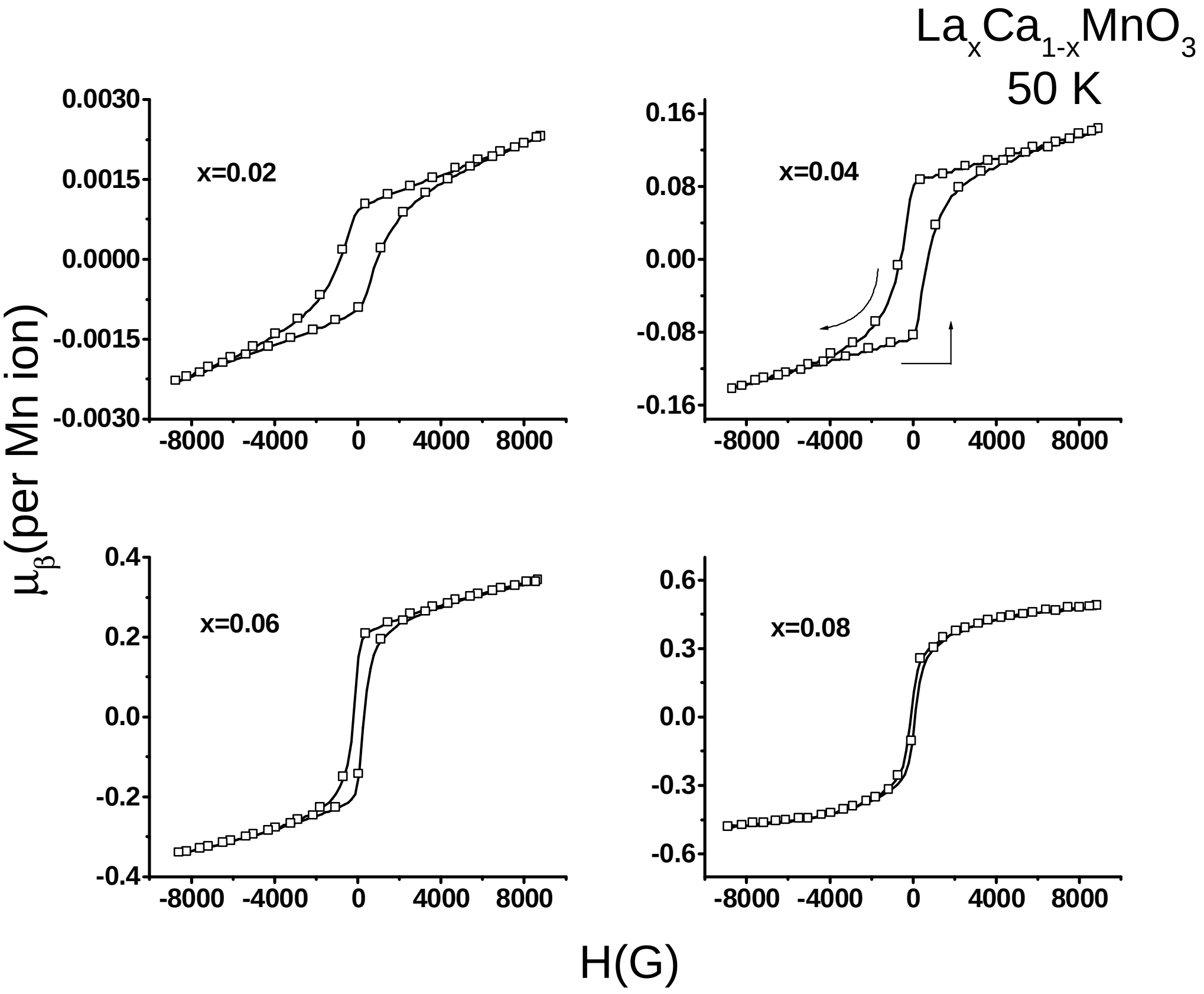

Fig.8 Magnetic hysteresis in $La_xCa_{1-x}MnO_3$ (x=0.02,0.04,0.06 and 0.08) at 50 K.

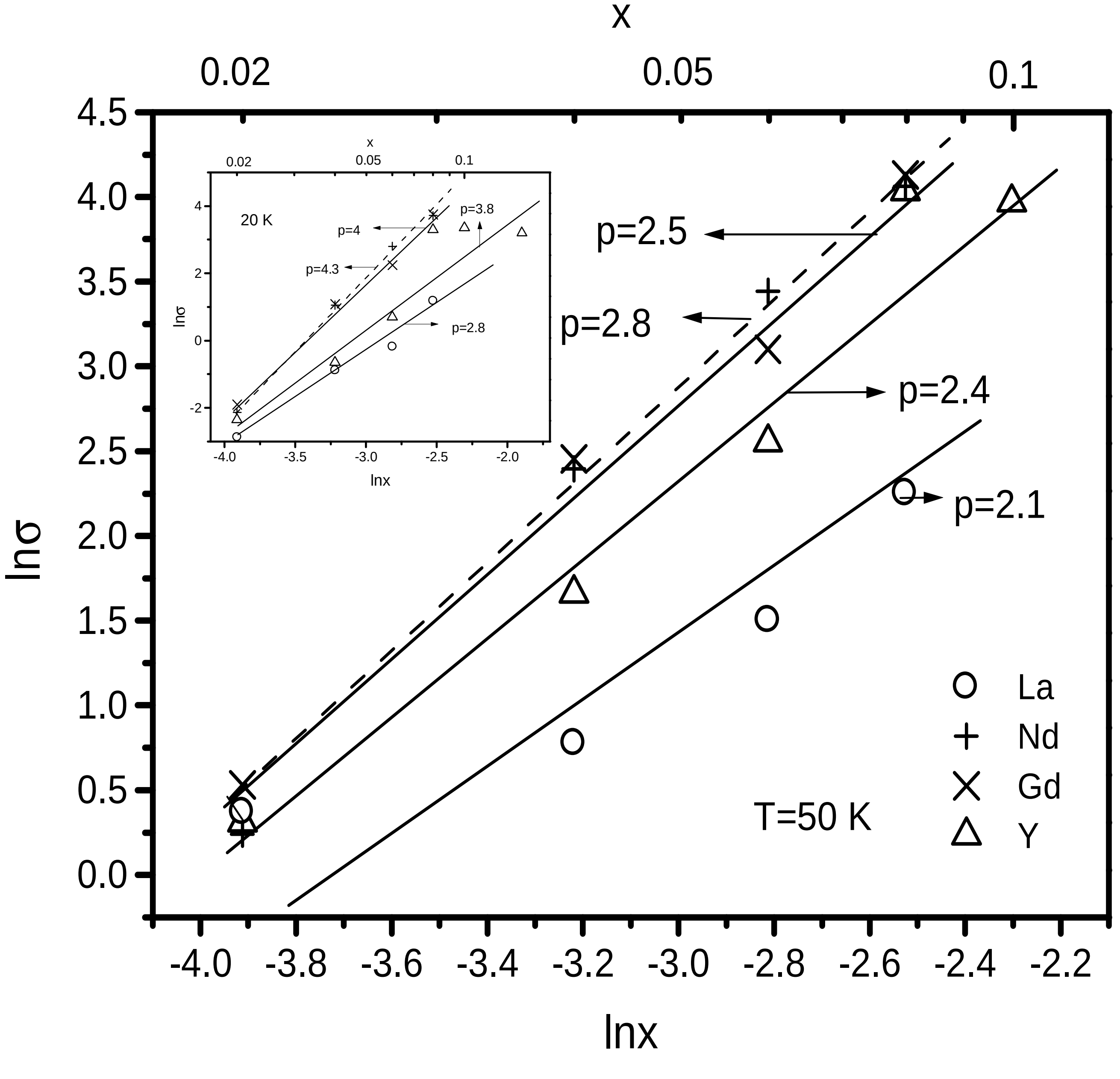

Fig.9 Conductivity data of $Ln_xCa_{1-x}MnO_3$ at 50 K fitted to the scaling law $\sigma \propto | x_c\text{-}x |^p$. Inset shows the conductivity data fitted to the equation at 20 K.